\documentclass[twoside]{article}

\usepackage{arxiv}

\usepackage[utf8]{inputenc}             % allow utf-8 input
\usepackage[T1]{fontenc}                % use 8-bit T1 fonts
\usepackage[numbers]{natbib}
\usepackage[linktocpage, colorlinks, linkcolor=blue, citecolor=blue]{hyperref}
\usepackage{url}                        % simple URL typesetting
\usepackage{booktabs}                   % professional-quality tables
\usepackage{amsfonts}                   % blackboard math symbols
\usepackage{nicefrac}                   % compact symbols for 1/2, etc.
\usepackage{microtype}                  % microtypography
\usepackage[nameinlink]{cleveref}       % smart cross-referencing
\usepackage{lipsum}                     % Can be removed after putting your text content
\usepackage{graphicx}
\usepackage{doi}
\usepackage{acro}

\usepackage{float}
\usepackage{graphicx}	% When possible, include images as *.eps format

\usepackage{scalerel}
\usepackage{tikz}
\usetikzlibrary{svg.path}

	\definecolor{orcidlogocol}{HTML}{A6CE39}
	\tikzset{
		orcidlogo/.pic={
			\fill[orcidlogocol] svg{M256,128c0,70.7-57.3,128-128,128C57.3,256,0,198.7,0,128C0,57.3,57.3,0,128,0C198.7,0,256,57.3,256,128z};
			\fill[white] svg{M86.3,186.2H70.9V79.1h15.4v48.4V186.2z}
			svg{M108.9,79.1h41.6c39.6,0,57,28.3,57,53.6c0,27.5-21.5,53.6-56.8,53.6h-41.8V79.1z M124.3,172.4h24.5c34.9,0,42.9-26.5,42.9-39.7c0-21.5-13.7-39.7-43.7-39.7h-23.7V172.4z}
			svg{M88.7,56.8c0,5.5-4.5,10.1-10.1,10.1c-5.6,0-10.1-4.6-10.1-10.1c0-5.6,4.5-10.1,10.1-10.1C84.2,46.7,88.7,51.3,88.7,56.8z};
		}
	}
	
	\newcommand\orcidicon[1]{\href{https://orcid.org/#1}{\mbox{\scalerel*{
					\begin{tikzpicture}[yscale=-1,transform shape]
						\pic{orcidlogo};
					\end{tikzpicture}
				}{|}}}}

\usepackage[obeyDraft]{todonotes}

\title{Separating Technological and Clinical Safety Assurance for Medical Devices}

\date{\today}
\author{    Spencer R. Deevy$^{\textsuperscript{\orcidicon{0000-0003-3331-1668}}}$\textsuperscript{1},
            Tiago de Moraes Machado$^{\textsuperscript{\orcidicon{0000-0002-6994-2181}}}$\textsuperscript{1},
            Amen Modhafar$^{\textsuperscript{\orcidicon{0000-0003-4887-2169}}}$\textsuperscript{2},
            Wesley O'Beirne$^{\textsuperscript{\orcidicon{0000-0002-6531-4057}}}$\textsuperscript{2},\\
            \textbf{Richard F. Paige$^{\textsuperscript{\orcidicon{0000-0002-1978-9852}}}$\textsuperscript{1},
            Alan Wassyng$^{\textsuperscript{\orcidicon{0000-0003-4614-3421}}}$\textsuperscript{1}}\\\\
	\textsuperscript{\textbf{1}} McMaster Centre for Software Certification, McMaster University, Hamilton, ON, Canada \\
	\textsuperscript{\textbf{2}} Arrayus Technologies Inc., Burlington, ON, Canada \\\\
	\textsuperscript{\textbf{1}} \texttt{\{deevys, machadotiagom, paigeri, wassyng\}@mcmaster.ca}\\
	\textsuperscript{\textbf{2}} \texttt{\{amodhafar, wobeirne\}@arrayus.ca}
}

\renewcommand{\undertitle}{}
\renewcommand{\shorttitle}{\small \textit{S. Deevy et al. - Separating Technological and Clinical Safety Assurance for Medical Devices}}

\hypersetup{
	pdftitle={Separating Technological and Clinical Safety Assurance for Medical Devices},
	pdfsubject={cs.SE, cs.SY, eess.SY},
	pdfauthor={Spencer Deevy, Tiago de Moraes Machado, Amen Modhafar, Wesley O'Beirne, Richard Paige, Alan Wassyng},
	pdfkeywords={Assurance Case, Separation of Concerns, Medical Devices, Safety-Critical Systems, Software-Intensive Systems, Safety, Certification, Medical Ultrasound, Focused Ultrasound, Clinical Effectiveness, Product Lines} 
}

\DeclareAcronym{mcscert}{short = McSCert, long = McMaster Centre for Software Certification}
\DeclareAcronym{ac}{short = AC, long = assurance case}
\DeclareAcronym{cae}{short = CAE, long = Claims-Arguments-Evidence}
\DeclareAcronym{cac}{short = CAC, long = Clinical Assurance Case}
\DeclareAcronym{tac}{short = TAC, long = Technological Assurance Case}
\DeclareAcronym{fda}{short = FDA, long = Food and Drug Administration}
\DeclareAcronym{fus}{short = FUS, long = focused ultrasound}
\DeclareAcronym{gsn}{short = GSN, long = Goal Structuring Notation}
\DeclareAcronym{mri}{short = MRI, long = magnetic resonance imaging}
\DeclareAcronym{mcpsos}{short = MCPSoS, long = Medical Cyber-Physical System of Systems}
\DeclareAcronym{omg}{short = OMG, long = Object Management Group}
\DeclareAcronym{pma}{short = PMA, long = Premarket Approval}
\DeclareAcronym{mrgfus}{short = MRgFUS, long = Magnetic Resonance-guided Focused Ultrasound}
\DeclareAcronym{sacm}{short = SACM, long = Structured Assurance Case Metamodel}
\DeclareAcronym{sos}{short = SoS, long = System of Systems}

\definecolor{blizzardblue}{rgb}{0.67, 0.9, 0.93}
\definecolor{celadon}{rgb}{0.67, 0.88, 0.69}
\definecolor{desertsand}{rgb}{0.93, 0.79, 0.69}
\definecolor{lilac}{rgb}{0.78, 0.64, 0.78}
\definecolor{paleaqua}{rgb}{0.74, 0.83, 0.9}

\begin{document}
\maketitle

\begin{abstract}
	The safety and clinical effectiveness of medical devices are closely associated with their specific use in clinical treatments. Assuring safety and the desired clinical effectiveness is challenging. Different people may react differently to the same treatment due to variability in their physiology and genetics. Thus, we need to consider the outputs and behaviour of the device itself as well as the effect of using the device to treat a wide variety of patients. High-intensity focused ultrasound systems and radiation therapy machines are examples of systems in which this is a primary concern. Conventional monolithic assurance cases are complex, and this complexity affects our ability to address these concerns adequately. Based on the principle of separation of concerns, we propose separating the assurance of the use of these types of systems in clinical treatments into two linked assurance cases. The first assurance case demonstrates the safety of the manufacturer's device independent of the clinical treatment. The second demonstrates the safety and clinical effectiveness of the device when it is used in a specific clinical treatment. We introduce the idea of these separate assurance cases, and describe briefly how they are separated and linked.
\end{abstract}

% keywords can be removed

%Sections
	% Introduction -------------------------------------------------------------
	\section{Introduction}
	\label{introduction}
	
	Modern medical devices are complex due to their intensive use of software. There are many kinds of medical devices. The interplay of treatment, medical indications, and inter-patient physiological variability introduces significant complexity with respect to safety when compared with other types of safety-critical system~\cite{hatcliff_2014}. There is also the tension between ensuring the safety of the device in terms of it not harming people, and the fact that not using the device may also lead to harm.
	
To achieve the intended safety and the intended clinical effectiveness, engineers must design, develop, manufacture and maintain their systems following the best safety engineering practices at hand. They also need to meet stringent functional safety standards such as IEC~62304~\cite{iec_62304_2006} and ISO~14971~\cite{iso_14971_2019}, as well as satisfy regulatory requirements.
	
Safety cases, a precursor of assurance cases, were introduced more than 50 years ago to help manufacturers document a structured, explicit argument that the system of 
interest is safe~\cite{bloomfield_2010}.
Modern assurance cases have the same intent, but are used to document that a system possesses  properties of concern, including but not limited to safety. Despite the benefits that assurance cases can bring to help develop safe systems as well as assure that they are safe, the adoption of assurance cases in medical device development varies by country and regulator. 

The \ac{mcscert} and Arrayus Technologies Inc. have been collaborating on the safety assurance of Arrayus's therapeutic \ac{fus} system. The device emits \ac{fus} energy waves to deliver precise treatment for several medical conditions, including uterine fibroids and pancreatic cancer. It uses an external \ac{mri} system for guidance of the treatment and monitoring of the patient. The combination of such non-invasive and non-ionizing technologies forms a system of systems commonly known as a \ac{mrgfus} system~\cite{bradley_2009}. One of the most difficult aspects of assuring safety of such a device is that there are so many variations in how different people react to the same treatment, even when using the same outputs from the medical device. To address this we suggest an approach for assuring safety and effectiveness of such devices. 
We propose separating the assurance into \textit{Technological Assurance} and \textit{Clinical Assurance}. Technological Assurance refers to a medical device system viewed solely as a machine that produces deterministic outputs given specific inputs. This is independent of the effect of these outputs on a patient during clinical treatment. Clinical Assurance refers to how those outputs from the machine affect patients within the clinical treatment.
This separation seems to be effective in reducing the complexity of the assurance.

%\input{Sections/Preliminaries}

% Proposed Assurance Case Separation for Medical Devices
\section{Proposed Assurance Case Separation for Medical Devices}
\label{sec:proposal}

\subsection{Goal Structuring Notation}
			In discussion related to assurance cases and in the structure of the assurance case figures, we have used Goal Structuring Notation (GSN)~\cite{gsn_community_standard_2011} with some minor changes in terminology. For example, we prefer to talk about \emph{claims} and \emph{evidence} rather than \emph{goals} and \emph{solutions}.
	
\subsection{The Monolithic Assurance Case}
 We typically create a single, comprehensive \ac{ac} for engineered systems in non-medical domains. In these assurance cases, safety is considered in relation to the overall behaviour of the system and the respective effects produced in a given environment. This strategy has been applied to medical devices as well, and has been effective for many of them, but is problematic for those that have to reckon with the fact that people come in many shapes and sizes: their bodies can respond differently to the same clinical treatment protocol. This variability, however, is not limited to patient physiology. It may also extend to the types of treatments a single medical device can perform and the different regions of the body that can be treated with that medical device. Therapeutic \ac{mrgfus} and radiation therapy machines are examples of how a complex system may be used to treat a wide variety of medical conditions, over several regions of the body. All of this makes it difficult to document a compelling assurance argument. The argument must demonstrate that the machine works as intended, delivers the correct outputs within a safe range, and if any output happens to be delivered outside of the intended location, it will not cause unacceptable harm to the patient, and will not be harmful to the environment -- all while achieving the desired physiological response for a particular patient.
	
\subsubsection*{Brief Example:}
\begin{figure}
		\centering
		\includegraphics[width=0.8\textwidth]{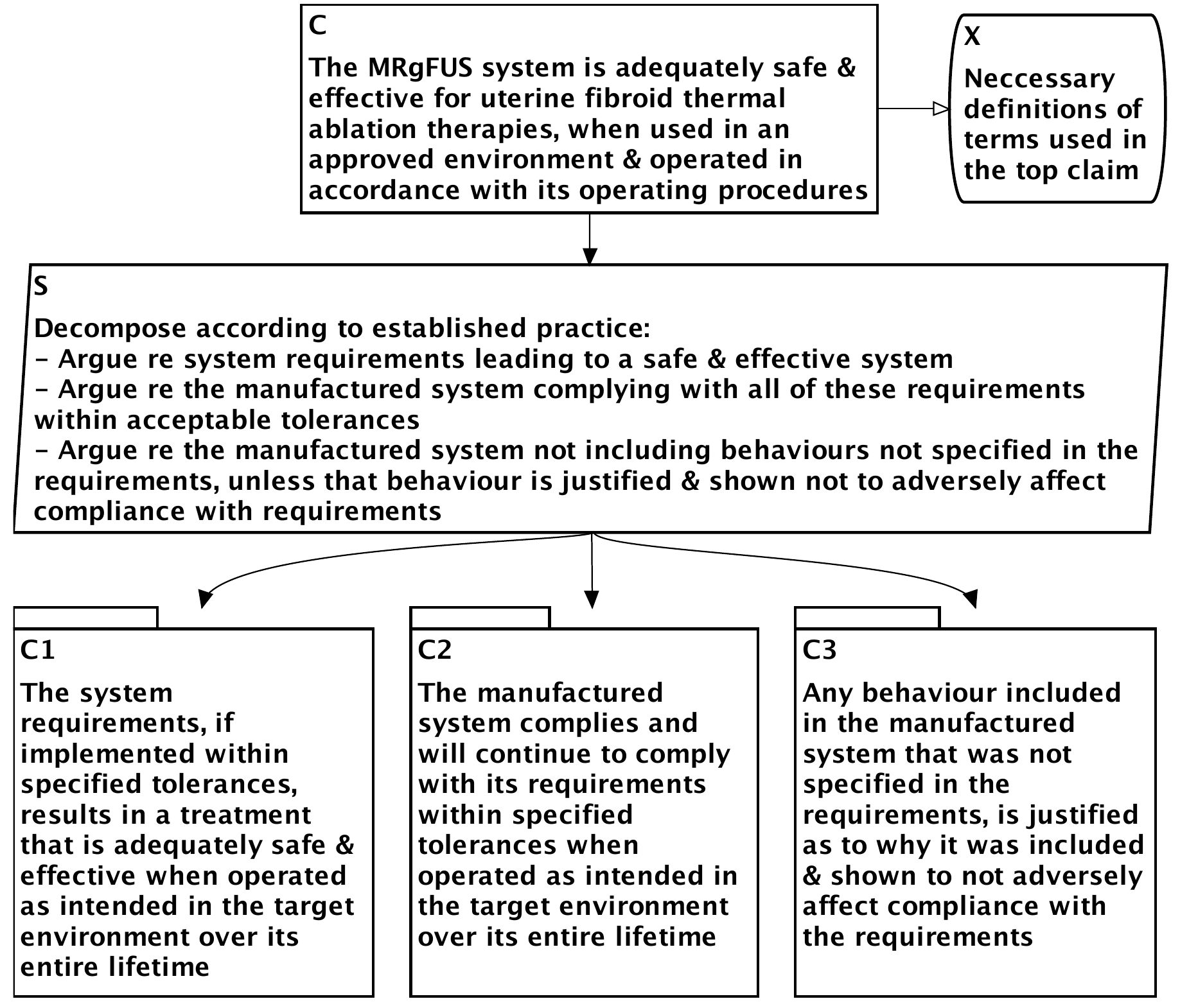}
		\caption{Top-level of a monolithic assurance case for a \ac{mrgfus} system}
		\label{monolithic figure}
	\end{figure}

	We now consider the top level of a monolithic assurance case for a \ac{fus} device that provides clinical treatment for uterine fibroids, achieved by thermally ablating problematic tissue. This is shown in \Cref{monolithic figure}, in which the top claim is shown along with its top-level GSN decomposition. The GSN components are labelled as follows: C indicates a \emph{Claim}; S indicates a \emph{Strategy}; %E indicates \emph{Evidence}; 
	and X indicates \emph{conteXt}. We have removed Assumptions and Justifications in the interest of saving space. The claims with the tabs on the top left edge are \emph{modules}. The lower levels of the argument are contained within those modules. The evidence that supports terminal claims in the argument are visible only in the content of those modules, and are not described in this paper.
	
\subsection{Separating Technological and Clinical Assurance Cases}
  Recently we realized an analogy between this situation and the complexity inherent in very sophisticated control systems that are also safety-critical. For years the nuclear industry has relied on separation of concerns to deal with such problems. Many countries have mandated the separation of control and safety. This results in much simpler safety systems that can then be built and certified to be safe, independent of what the control system does. This separation is enforced in the system itself. It occurred to us that for some medical systems, separation of concerns to control complexity could be applied to the assurance case itself.
  
We first need to define what we mean by \textit{Technological effects} and \textit{Clinical effects}.

\subsubsection{Technological effects}
            \label{technological effects}
				When considering the \textit{technological effects} of the medical device, we consider the device solely as a machine that produces deterministic outputs given specific inputs. It does not include how the output of the medical device affects patients.  

\subsubsection{Clinical effects}
            \label{clinical effects}
			    The \textit{clinical effects} of the device refers to the physiological response of a human patient to the use of the medical device and its operating procedures during a specific clinical treatment. This is meant to cope with the fact that different people can react differently to the exact same treatment. 

\subsubsection{Splitting the Argument Based on Technological Effects and Clinical Effects}
 Instead of constructing a monolithic assurance case, we propose splitting the argument of safety and effectiveness for certain medical devices into two linked assurance cases: one based on the \emph{``Technological effects''} as defined above, and another based on the \emph{``Clinical effects''}, also defined above. The former presents the argument pertaining to the safety and the effectiveness of the medical device and the therapy-agnostic operating procedures in relation to the medical device's ability to deliver its promised behaviour independent of any clinical context.
 
  The latter presents the argument pertaining to the safety and the effectiveness of the medical device and therapy-specific operating procedures/treatment plans in relation to achieving the intended biological/physiological response required to treat a particular medical indication. Overall assurance of the medical device used for a specific therapy is obtained by the combination of two linked \acp{ac}. The final assurance is documented in the \ac{ac} that is focused on the clinical effects. We call this \ac{ac} the \ac{cac}). The \ac{cac} is dependent on assurance provided by the \ac{ac} that focuses on the technological effects. We call this \ac{ac} the (\ac{tac}).
	
We now compare the monolithic \ac{ac} example with the proposed separation of assurance cases in the following sections. 

	% Technological Assurance Case
	\subsubsection{Technological Assurance Case}
	\label{tac}
	
\begin{figure}[H]
			\centering
			\includegraphics[width=0.9\textwidth]{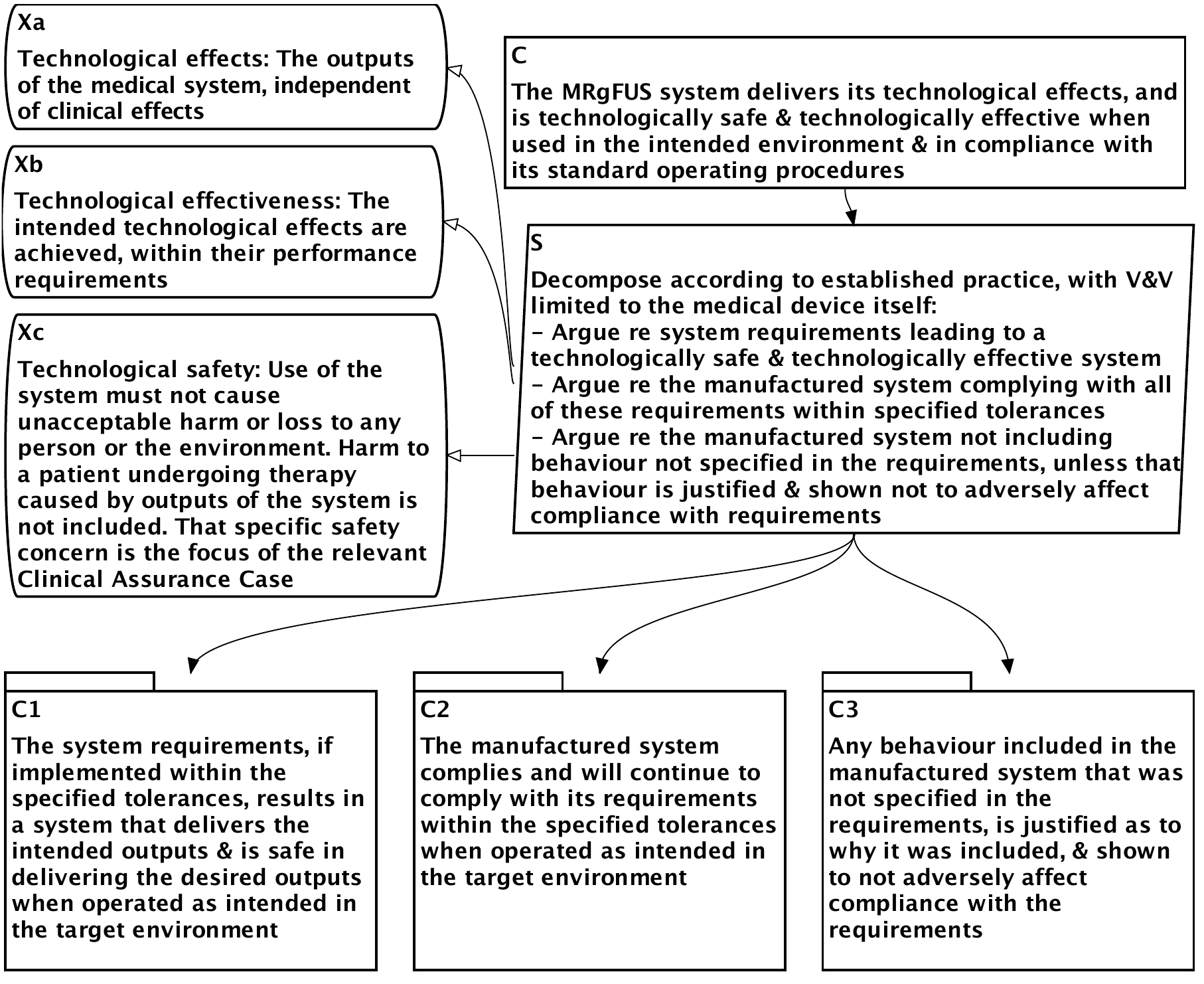}
			\caption{Top-level of the \ac{tac} for a \ac{mrgfus} system}
			\label{tac figure}
		\end{figure}

	The top claim and top-level decomposition of the \emph{\ac{tac}} is shown in \Cref{tac figure}. 
		
	This is slightly different from the monolithic version in that it disregards the clinical application and treats the safety of the system according to its capability in delivering the technological effects, or outputs, independent of any clinical effect. The device must be able to focus and deliver ultrasonic energy to a particular location in space from its ultrasonic transducers within the required specifications. Such performance-based properties of the technological effects are referred to as \textit{technological effectiveness}. The medical device must also handle safety concerns that affect everyone using the system. These safety concerns are identified through hazard analyses and deal with the intended behaviour of the system as well as how the machine interfaces with its environment and interacts with other medical devices. These safety concerns are also independent of any specific clinical effects. The safety-based properties of the technological effects are referred to as \textit{technological safety}.
	
Our standard practice in structuring the argument is described in the Strategy, S. The context components Xa, Xb and Xc define technological effects, and the effectiveness and safety properties associated with the technological effects, and all three of these terms reinforce that the effect of the outputs of the medical device on the patient is not considered within this argument.  Those safety and effectiveness concerns are to be separated out and dealt with in the \ac{cac}.

	% Clinical Assurance Case
	\subsubsection{Clinical Assurance Case}
	\label{cac}

\begin{figure}
			\centering
			\includegraphics[width=0.9\textwidth]{"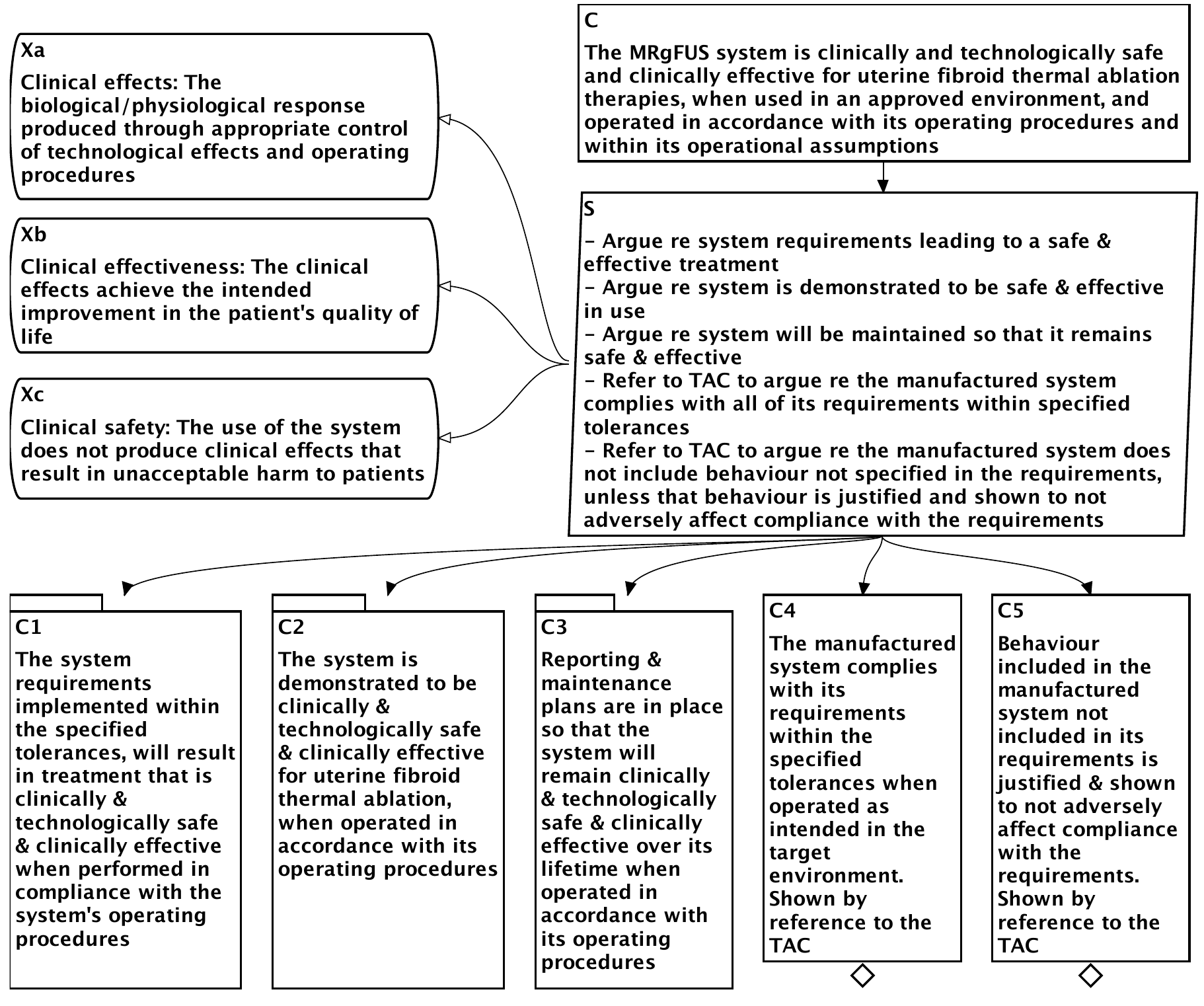"}
			\caption{Top-level of the \ac{cac} for a \ac{mrgfus} system}
			\label{cac figure}
		\end{figure}

    The second assurance case we have called the \emph{\ac{cac}} and its top-level is shown in \Cref{cac figure}. We use the same notation in the \ac{cac} as we did in the \ac{tac} in \Cref{tac figure}. As one can see, the top claim in the CAC is different from the top claim in the TAC, since the meaning of safety and effectiveness for the CAC now addresses the intended clinical effects relevant to the clinical application of the medical device. 
    
We use a modified version of our standard practice in the strategy S, where the first subclaim involves showing that the system requirements result in a safe and effective treatment (clinical effects), where we build off of the work that the \ac{tac} demonstrated to show that the system is clinically sufficient for a given treatment. Xa, Xb, and Xc give the corresponding definitions for clinical effects, \textit{clinical effectiveness}, and \textit{clinical safety}. These differ from their corresponding definitions in the \ac{tac} in that they are now in the context of a clinical setting, and more importantly, how we use the medical device to achieve the intended physiological response in a patient safely and effectively.

\subsubsection{How the \ac{tac} and \ac{cac} are Linked}
We can see in \Cref{cac figure} that claims C4 and C5 are similar to claims C2 and C3 in the monolithic assurance case, and are dealt with by reference to the associated \ac{tac}. They do not have to be argued in the \ac{cac}! Clearly the \ac{cac} is dependent on its associated \ac{tac} in that the safety of the machine itself in delivering its functional outputs is documented in the \ac{tac}. This implies that the outputs of the medical system required for a clinical treatment must be documented explicitly in the \ac{cac}, and then verified as provided by the system as documented in the \ac{tac}. In general, the \ac{cac} may reference any items in the \ac{tac}. However, it is crucial that there are no references from the \ac{tac} to any dependent \ac{cac}. The diamonds below the claims C4 and C5 are GSN symbols to indicate that the claims are not further developed in the \ac{cac}. The required information is documented in context nodes that support claims C4 and C5.

%\input{Sections/Benefits}	

%\input{Sections/RelatedWork}

% Conclusion
\section{Conclusion}
\label{conclusion}

	In practice, we can assure the safety of the medical system independent of its clinical effects in a \ac{tac}, and the safety of its clinical application in an associated \ac{cac}. The assurance cases are linked, and both are needed to provide full assurance for a particular treatment. (There is always the option of combining multiple clinical treatments in a single \ac{cac}, or developing separate \acp{cac} for different clinical applications.) Every \ac{cac} builds on the assurance documented in its associated \ac{tac}. As long as the device documented in the \ac{tac} has the capability to perform the clinical application, the \ac{tac} does not have to be modified. This presents the basic idea behind the separation of the \ac{tac} and \ac{cac}. 

	We have shown that separation of concerns can be used in assurance cases to reduce the complexity of demonstrating safety and effectiveness for software-intensive medical systems, such as the \ac{mrgfus}. By separating the demonstration of system safety in producing the intended deterministic machine output independent of clinical safety,  we believe we can significantly reduce the overall complexity of the safety assurance argument.

 It is common to find that a particular medical device is used for different clinical procedures as is the case for the \ac{mrgfus}. If we assure the technological safety and effectiveness of the device independent of its clinical effects, then it raises the possibility that a single \ac{tac} could be linked with multiple \acp{cac} to assure the safety and effectiveness of the device used in multiple clinical treatments.

% Bibliography -------------------------------------------------------------
%\bibliographystyle{unsrtnat}
\bibliographystyle{unsrt} % use this for having a numerical reference style
\bibliography{citations}

\end{document}